\documentclass{iaus}
\usepackage{graphicx}

\def\aj{AJ}%
\def\araa{ARA\&A}%
\def\apj{ApJ}%
\def\apjl{ApJ}%
\def\apjs{ApJS}%
\def\aap{A\&A}%
\def\mnras{MNRAS}%
\def\nat{Nature}%
\title[Massive Star Formation] %% give here short title %%
{Clustered Massive Star Formation in Molecular Clouds}

\author[Jonathan C. Tan]   %% give here short author list %%
{Jonathan C. Tan}%$^1$%
\affiliation{$^1$Inst. of Astronomy, Dept. of Physics, ETH Z\"urich, 8093 Z\"urich, Switzerland; jt@phys.ethz.ch} 

\pubyear{2005}
\volume{227}  %% insert here IAU Symposium No.
\pagerange{???--???}
\date{?? and in revised form ??}
\jname{Massive Star Formation: A Crossroads in Astrophysics}
\editors{R. Cesaroni, E. Churchwell, M. Felli, and C. M. Walmsley., eds.}
\begin{document}

\maketitle

\begin{abstract}
I review some important questions in the field of massive star
formation: What are the initial conditions for proto star
clusters and how do they arise? What are the initial conditions for
individual massive star formation within star clusters?  How do
massive protostars accumulate their mass? I compare the Turbulent Core
Model (McKee \& Tan 2003) to several nearby regions, including Orion
KL. Here I also discuss the origin of BN's high proper motion.
\end{abstract}

\firstsection % if your document starts with a section,
\section{Introduction}

Throughout cosmic history, massive stars have been the principal
drivers of galactic evolution, energizing and enriching gas,
regulating its collapse into further stellar generations or even
ejecting it into intergalactic space. The importance of massive stars
is extended by the realization that a substantial fraction of all
stars, and perhaps planetary systems, form in star clusters (Lada \&
Lada 2003), where proximity amplifies the influence of massive star
feedback. The fueling of supermassive black holes, from the Milky
Way's relatively small example to the giants powering quasars, is also
likely to be influenced by massive star formation (Goodman \& Tan
2004; Thompson et al. 2005).

Given this importance, our lack of understanding of massive star
formation is shocking. For example, there has been no clear consensus
on even the basic formation mechanism, whether it be collapse from
approximately stellar mass gas cores via accretion disks
(e.g. McLaughlin \& Pudritz 1997; Osorio et al. 1999; Yorke \&
Sonnhalter 2002; McKee \& Tan 2003, hereafter MT03), competitive
accretion of ambient cluster gas (Bonnell et al. 2001; Bonnell, Vine,
\& Bate 2004; Schmeja \& Klessen 2004), or, most radically, direct
stellar collisions (Bonnell, Bate, \& Zinnecker 1998). Even amongst
the more conventional core models, a vast range of parameters, such as
formation timescale or accretion rate, have been
discussed. Present-day massive stars appear to form exclusively within
star clusters (de Wit et al. 2005), but the timescale for star cluster
formation is debated --- does it take few (Elmegreen 2000) or many
(Tan 2004a) dynamical times?  The influence of feedback in setting
both the stellar initial mass function, including its upper limit, and
the efficiency of star formation in clusters, is uncertain.
Observationally, the closest example of a massive protostar (i.e. a
star undergoing active growth by accretion), in the Orion
Kleinmann-Low (KL) region, is so confusing that it has been used in
support of all the different models (Tan 2004b; Bally \& Zinnecker
2005). Even amongst proponents of conventional models with accretion
disks, the orientation of this disk axis is debated between two models
that are orthogonal to each other (Greenhill et al. 2003; Tan 2004b).

The reasons for this confusion are, on the theoretical side, the range
of scales that must be analyzed or simulated, the complicated physics
exhibited by a self-gravitating, magnetized, turbulent, optically
thick, chemically-evolving medium, and the large and uncertain
parameter space and boundary conditions for the models and
simulations. Observationally, the problems include the relative rarity
of massive protostars, their typically large distance from us, their
high obscuration, and the crowding of their environments.

In this review, I focus first on the initial conditions of star
clusters (\S2) and then of massive stars and their mode of
accretion(\S3). My basic conclusion is that a model of formation
involving the collapse of gas clumps and cores in approximate pressure
equilibrium with their surroundings and in approximate virial
equilibrium, can explain both star cluster and massive star
formation. I give some examples of how the model applies to
individual, well-studied regions of massive star formation (\S4),
focusing particularly on Orion KL. Here I also give an update on the
discussion of the origin of the Becklin-Neugebauer (BN) object, which
is prominent in this region. Due to lack of space, I do not discuss
feedback in star clusters or the timescale
of star cluster formation.

\section{The Initial Conditions for Star Cluster Formation}

Figure~1 reviews the masses, $M$, and mean surface densities,
$\Sigma=M/(\pi R^2)$, of local star clusters and interstellar gas
clouds.  For convenience $\Sigma = 1\:{\rm g\:cm^{-2}}$ corresponds to
$4800\:M_\odot\:{\rm pc^{-2}}$, $N_{\rm H}=4.3\times 10^{23}\:{\rm
cm^{-2}}$ and $A_V=200$~mag, for the local gas to dust ratio.
Contours of constant radial size, $R$, and H number density,
$n_{\rm H} = \rho/\mu = 3M / (4\pi R^3 \mu)$ where $\mu=2.34\times
10^{-24}\:{\rm g}$ is the mean mass per H, are indicated. Density
contours correspond to free-fall times, $t_{\rm
ff}=\sqrt{3\pi/(32G\rho)}=1.38\times 10^{6} (n_{\rm H}/10^3\:{\rm
cm^{-3}})^{-1/2}\:{\rm yr}$.
%For a virialized cloud with virial
%parameter $\alpha_{\rm vir}\simeq 1$ (Bertoldi \& McKee 1992) the
%signal crossing or dynamical timescale is $t_{\rm dyn}=2.0t_{\rm ff}$.

The presence of molecules allows gas to cool to low temperatures,
$\sim 10\:{\rm K}$, effectively removing thermal pressure support. To
survive the destructive local interstellar FUV radiation requires a
total column of $N_{\rm H}=(0.4,2.8)\times 10^{21}\:{\rm cm^{-2}}$ for
$\rm H_2$ and CO, respectively (McKee 1999). Giant molecular clouds
(GMCs) have a constant column of $N_{\rm H}\simeq(1.5\pm0.3)\times
10^{22}\:{\rm cm^{-2}}$ and typical masses $\sim 10^5-10^6\:M_\odot$
(Solomon et al. 1987). A sample of local ($d\lesssim 3$~kpc)
infrared-dark clouds (IRDCs) have masses of a few$\times10^3 -
10^4\:M_\odot$ and $\Sigma\sim 0.1\:{\rm g\:cm^{-2}}$ (Kirkland \&
Tan, in prep.), about 3 times the GMC mean. Several IRDCs with higher
surface densities, $\sim 0.5-50\:{\rm g\:cm^{-2}}$, have been reported
by Carey et al. (1998). Star-forming clumps observed in the sub-mm by
Mueller et al. (2002) have similar masses, but $\Sigma\sim 0.1-1\:{\rm
g\:cm^{-2}}$.  More revealed systems, e.g. the Orion Nebula Cluster
(ONC), have similar properties. These considerations suggest that
IRDCs, forming from GMCs, are the initial conditions for star
clusters. It also appears that large masses can accumulate in IRDCs
before the onset of significant star formation.

Galactic star clusters more massive than $\sim 10^4\:M_\odot$ are
quite rare. The Arches and Quintuplet clusters (e.g.  Kim et al. 2000)
are examples in the Galactic center region. The most massive young
clusters, so-called super star clusters, are often found in starburst
environments, e.g. the Antennae galaxies (Mengel et al. 2001), and in
some dwarf galaxies, e.g.  NGC~5253 (Turner et al. 2000) and NGC~1569
(Gilbert \& Graham 2003).

All these high-mass star-forming regions appear to have a density of
$n_{\rm H} \sim 2\times 10^5\:{\rm cm^{-3}}$,
%, corresponding to $t_{\rm ff}\sim 1 \times 10^5\:{\rm yr}$. 
about that at which the cooling rate is a maximum (Larson 2005), and thus
gravitational collapse is easiest. A spherical self-gravitating cloud
in hydrostatic equilibrium with mean surface density $\Sigma$ and
density profile $\rho\propto r^{-k_\rho}$ with $k_\rho=1.5$, similar
to observed clumps, has a mean pressure of $4.3 \times
10^8\Sigma^2\:{\rm K~cm^{-3}}$ (MT03). Massive stars and
star clusters form under pressures $\gtrsim 3\times 10^7\:{\rm
K~cm^{-3}}$, much higher than that of the local diffuse ISM,
i.e. $2.8\times 10^4\:{\rm K~cm^{-3}}$ (Boulares \& Cox 1990).

\begin{figure}
\includegraphics[height=5.2in,width=5.2in,angle=0]{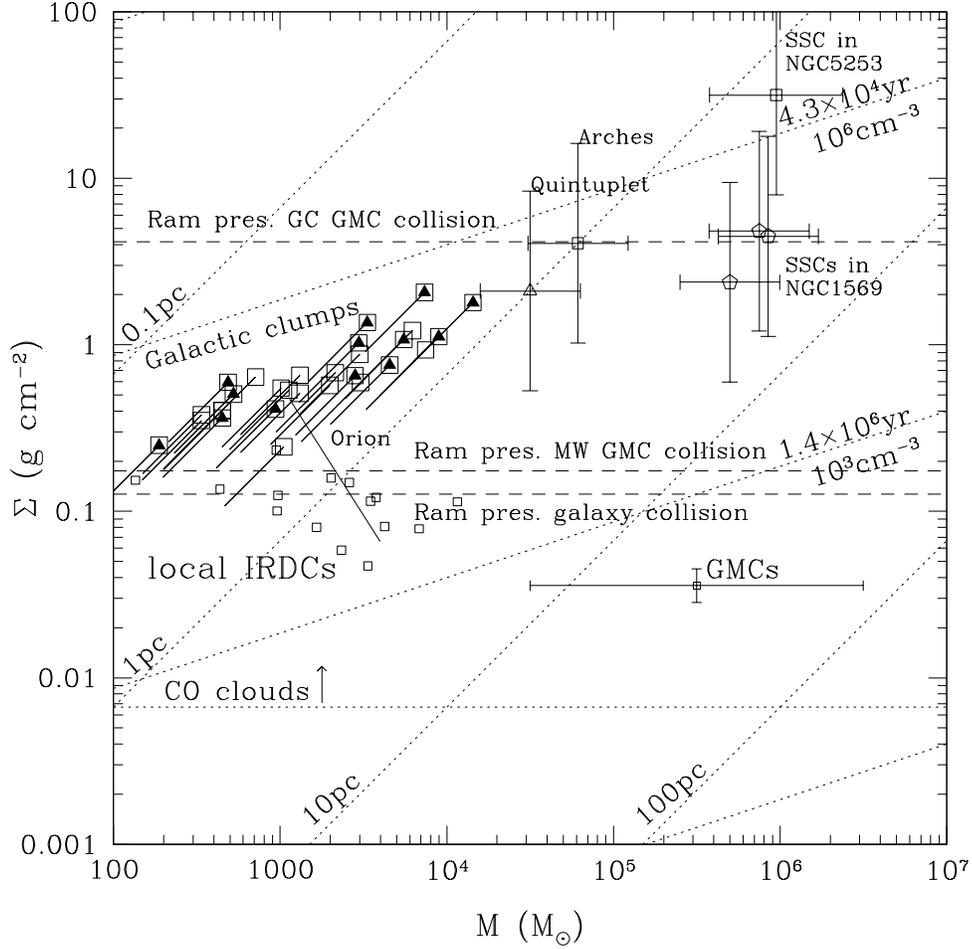}
  \caption{Surface density, $\Sigma$, versus mass, $M$, for star
  clusters and interstellar clouds. Contours of constant radius , $R$,
  and hydrogen number density, $n_{\rm H}$, or free-fall timescale,
  $t_{\rm ff}$, are shown with dotted lines. The minimum $\Sigma$ for
  CO clouds in the local Galactic FUV radiation field is shown, as are
  typical GMC parameters. Local infrared dark clouds are shown by
  small open squares (Kirkland \& Tan, in prep.). Large open squares
  are star-forming clumps (Mueller et al. 2002): a triangle indicates
  the clump contains an HII region; the diagonal lines show the effect
  of uncertain dust opacities on the mass estimate. The Orion Nebula
  Cluster, assuming equal mass in gas and stars, is shown by the solid
  line, which traces conditions from the inner to the outer parts of
  the cluster. Several more massive clusters are also indicated (see
  \S2). Ram pressures from a typical Galactic GMC-GMC collision
  ($v=10\:{\rm km\:s^{-1}}$, $n_{\rm H}=500\:{\rm cm^{-3}}$), a
  Galactic Center GMC-GMC collision ($v=40\:{\rm km\:s^{-1}}$, $n_{\rm
  H}=2.5\times 10^4\:{\rm cm^{-3}}$), and in the diffuse ISM of a
  galaxy-galaxy collision ($v=200\:{\rm km\:s^{-1}}$, $n_{\rm
  H}=1\:{\rm cm^{-3}}$) are indicated by horizontal dashed lines: note
  pressures are converted to an equivalent $\Sigma$ that would
  create the same mean pressure in a spherical, self-gravitating cloud,
  i.e. $P/k=4.3 \times 10^8\Sigma^2\:{\rm K~cm^{-3}}$ (MT03).  }
\label{fig:1}
\end{figure}

What causes a particular region of a GMC to form a star cluster? From
Figure~1 we see that the surface density, pressures, and volume
densities must increase by at least factors of 10, 100, and 1000,
respectively. This occurs in only a small part of the GMC: typically
only $\sim$1\% of the mass is involved. Models for the cause of star
formation can be divided into two groups: quiescent and triggered. In
the former, star formation occurs in the densest, most unstable clumps
of the GMC, and these form out of the general gravitational
contraction of the entire cloud. This process may be regulated by the
decay of turbulence, ambipolar (Mouschovias 1996) or turbulent
diffusion of magnetic flux, or heating and ionization (McKee 1989)
from newly-formed star clusters.

In triggered models the star-forming clumps are created by compression
of parts of the GMC by external causes, such as: cloud collisions
(Scoville et al. 1986; Tan 2000); convergent turbulent flows (e.g. Mac
Low \& Klessen 2004); or feedback from young stars with ionization
(Elmegreen \& Lada 1977; Deharveng et al. 2005), stellar winds (e.g.
Whitworth \& Francis 2002), protostellar outflows, radiation pressure,
and supernovae (e.g. Palous et al. 1994).  Elmegreen (2004) noted
that compressions resulting from most forms of stellar feedback
are probably only efficient within particular GMCs or GMC complexes,
i.e. young stars forming in one GMC are unlikely to trigger star
formation in another. Oey et al. (2005) claim the age sequence of 3
regions of the W3/W4 complex is evidence for triggered star formation
over an approximately $100$~pc scale region.

Considering cloud collisions or convergent flows, the typical ram
pressures, $P = \rho v^2 \rightarrow 1.73\times10^7 (n_{\rm H}/1000
{\rm cm^{-3}}) (v/10 {\rm km\:s^{-1}})^2\:{\rm K\:cm^{-3}}$, created
in a variety of interactions are shown in Fig.~1. GMC collisions could
compress local regions to pressures relevant to IRDCs. Pre-existing
dense, stable clumps that become enveloped by a newly over-pressured
region would then start to collapse to form star clusters. Such a
mechanism predicts a high degree of correlation in the locations
of young star clusters within GMCs.

\section{The Initial Conditions and Mode of Accretion of Massive Stars}

%describe the debate
Star clusters are born from turbulent gas, i.e. having velocity
dispersions much greater than thermal. A basic question is how
individual stars, particularly massive stars, form in this
environment. Do they grow inside quasi-equilibrium gas {\it cores}
that collapse via accretion disks with relatively stable orientations?
In this scenario (e.g. Shu, Adams, \& Lizano 1987, MT03) the initial
mass of the core helps to determine the final mass of the star, modulo
the effects of protostellar feedback.  Alternative models involving
competitive Bondi-Hoyle accretion (Bonnell et al. 2001; Bonnell et
al. 2004; Schmeja \& Klessen 2004) and direct stellar collisions
(Bonnell, Bate, \& Zinnecker 1998) have been proposed. These models
have been particularly motivated for massive star formation since this
occurs in the most crowded regions, radiation pressure feedback from
massive stars on dust grains can cause problems for standard accretion
scenarios, and the Jeans mass in these high pressure, high density
regions is only a fraction of a solar mass.

\subsection{Formation of Cores}

%describe core formation and observed starless cores
First consider the formation of cores from a turbulent medium.
Klessen et al. (2005) find that a substantial fraction of ``cores''
identified in their nonmagnetic SPH simulations of supersonic
turbulence appear quiescent (line widths $\leq$ thermal) and
coherent (line widths roughly independent of positional
offset from core center), but are in fact dynamic and transient.
They argue that the inference of hydrostatic equilibrium,
e.g. from radial profiles appearing similar to Bonnor-Ebert profiles
(Alves, Lada, \& Lada 2001), is not necessarily valid, since such
profiles are also possible for dynamically evolving cores. However, it
is not clear if these artificial dynamic cores are consistent with the
observations of Walsh et al. (2004), who find very small ($\lesssim
0.1\:{\rm km\:s^{-1}}$) velocity differences between the line centers
of high ($n_{\rm H} \sim 4\times 10^{5}\:{\rm cm^{-3}}$) and low
($n_{\rm H} \sim 2\times 10^{3}\:{\rm cm^{-3}}$) density traces of
starless cores, i.e. these cores are not moving with respect to
their envelopes.

%describe how magnetic fields affect results
The numerical simulations described above are
nonmagnetic. V\'azquez-Semadeni et al. (2005) studied core formation
from turbulent, magnetized gas. They find in their periodic, fixed
grid, isothermal, ideal MHD, driven turbulence simulations, that:
magnetic fields reduce the probability of core formation; in the
magnetically subcritical run, a bound core forms that lasts $\sim 5
t_{\rm ff}$ (defined at densities $\sim 50$ times the simulation
mean), which is long enough for ambipolar diffusion to affect the
dynamics; in the moderately supercritical case, where magnetic fields
are relatively weaker, bound cores form and then undergo runaway
collapse over about $2t_{\rm ff}$, defined at the core's mean
density. These results suggest that the initial conditions for star
formation are bound cores, and that the stronger the magnetic field,
the more chance cores have to attain a quasi-equilibrium
structure. The marginally critical case is probably most relevant if
star-forming clumps evolve from regions of GMCs that gradually lose
magnetic support. Crutcher's (1999) observations of the magnetic field
strength in dense regions of GMCs imply that these regions are only
marginally supercritical and that magnetic fields are dynamically
important.

%the mass function of cores
Magnetic fields are likely to affect the mass function of cores. One
argument against massive star formation from massive cores has been
that the thermal Jeans mass in the high pressure, high density regions
associated with massive star formation is very small. However this is
irrelevant if massive cores derive most of their pressure support from
either magnetic fields or turbulent motions. Observations suggest that
the mass function of cores is fairly similar, within large
uncertainties, to that of stars and that there are some massive
pre-stellar cores (e.g. Beuther \& Schilke 2004).

\subsection{Accretion to Stars}

%describe simulations that attempt to follow star formation, i.e. with sink particles.
It is computationally expensive to follow gravitational collapse to
the high densities and short timescales associated with protostars and
their accretion disks. A common numerical technique is to introduce
sink particles in bound regions of high density, which can then
accrete gas from their surroundings (Bate, Bonnell, \& Price 1995).

Bonnell et al. (2004) modeled star cluster formation with SPH,
isothermal, non-periodic, no feedback, nonmagnetic simulations, with
initially turbulent gas (no later driving) and sink particles about to
undergo global collapse.  Stars gained mass by competitive
accretion. The most massive star acquired gas that was initially very
widely distributed. Dobbs, Bonnell, \& Clark (2005) found that a
massive turbulent core, such as envisaged by MT03, fragments into many
smaller cores if the equation of state is isothermal.  However, their
non-isothermal model suffered much less fragmentation.
%Schmeja \& Klessen (2004) simulated
%star cluster formation with SPH simulations with periodic boundaries,
%driven turbulence, no magnetic fields, no feedback, sink particle
%diameters of 560~AU, and an isothermal equation of state, finding
%highly variable accretion rates for their protostars.

%problem of competitive accretion to sink particles.
We have seen that SPH simulations, by lacking magnetic fields,
probably do not accurately model the fragmentation process of real
star-forming regions, particularly with regard to core formation.
Another difficulty is in resolving the accretion by sink particles of
gas with vorticity (Krumholz, McKee, \& Klein 2005), with SPH
simulations tending to greatly overestimate the accretion
rate. Stellar feedback should also reduce this accretion rate,
particularly to massive stars (e.g. Edgar \& Clarke 2004). Thus the
importance of competitive accretion may be grossly overestimated in
current SPH simulations.

\subsection{Assumptions and Predictions of the Turbulent Core Model}

%Relation of all this to McKee-Tan model
McKee \& Tan (2002; 2003, MT03) modeled massive star formation by
assuming an initial condition that is a marginally unstable, massive,
turbulent core in approximate pressure equilibrium with the
surrounding protocluster medium, i.e. the star-forming clump.  To
derive the pressure in the clump, it was also assumed to be in
approximate hydrostatic equilibrium so that $P\sim G\Sigma^2$. This
pressure sets the overall density normalization of each core and thus
its collapse time and accretion rate.

The core is assumed to collapse via an accretion disk to form a single
star or binary. The assumption of collapse in isolation is
approximate: MT03 estimate that during the collapse the core interacts
with a mass of ambient gas similar to its initial mass, although
not all of this will become bound to the core.  The core density
structure adopted by MT03 is $\rho \propto r^{-k_\rho}$, with
$k_\rho=1.5$ set from observations. This choice affects the evolution
of the accretion rate: $k_\rho<2$ implies
accretion rates accelerate. However, this is a secondary effect
compared to the overall normalization of the accretion rate that is
set by the external pressure. Also, since much of the pressure support
is nonthermal with significant contributions from turbulent motions,
one does not expect a smooth density distribution in the collapsing
core, and the accretion rate can show large variations about the
mean.

%The assumption of approximate pressure equilibrium in the protocluster
%requires star formation to occur over at least several dynamical
%timescales, and this is examined in the next section. The basic
%picture of star cluster formation then involves: a turbulent,
%self-gravitating gas clump in which bound cores occasionally form
%(most gas at any given time is not in bound, unstable cores); a core
%mass function fairly similar to stars, i.e. massive cores form but are
%rare; an approximate equilibrium of cores with their surroundings; the
%collapse of cores quite rapidly in one or two free-fall timescales to
%form stars or binaries; the orbiting of newly-formed stars in the still
%star-forming clump, but negligible growth via competitive accretion.

Some key predictions are the properties of the cores and accretion
disks of massive stars. The core size is $R_{\rm core} \simeq 0.06
(M_{\rm core}/60M_\odot)^{1/2}\Sigma^{-1/2}\:{\rm pc}$. Note an
allowance has been made for massive cores tending to be near the
centers of clumps, where pressures are about twice the mean (MT03).
These cores have relatively small cross-sections for close
interactions with other stars. The accretion rate to the star, via a
disk, is $\dot{m}_* = 4.6\times 10^{-4} f_*^{1/2}
M_{60}^{3/4}\Sigma^{3/4}\:M_\odot\:{\rm yr}^{-1}$, where $f_*$ is the
ratio of $m_*$ to the final stellar mass and a 50\% formation
efficiency due to protostellar outflows is assumed. The collapse time,
$1.3\times 10^5 M_{60}^{1/4} \Sigma^{-3/4}\:{\rm yr}$, is short and
quite insensitive to $M$, allowing coeval stochastic high and low mass star
formation in a cluster, that might take $\gtrsim 1$~Myr to build up.
The disk size is $R_{\rm disk}=1200 (\beta/0.02) (f_* M_{60})^{1/2}
\Sigma^{-1/2}{\rm AU}$, where $\beta$ is the initial ratio of
rotational to gravitational energy of the core, and the normalization
is taken from typical low-mass cores (Goodman et al. 1993), although
there is quite a large dispersion about this value. These estimates
allow quantitative models of the protostellar evolution, disk
structure and outflow intensity, which can then be compared to
observed systems.

\section{Comparison of the Turbulent Core Model to Observed Regions}\label{sec:comparison}

\begin{figure}
\includegraphics[height=3.75in,width=5in,angle=0]{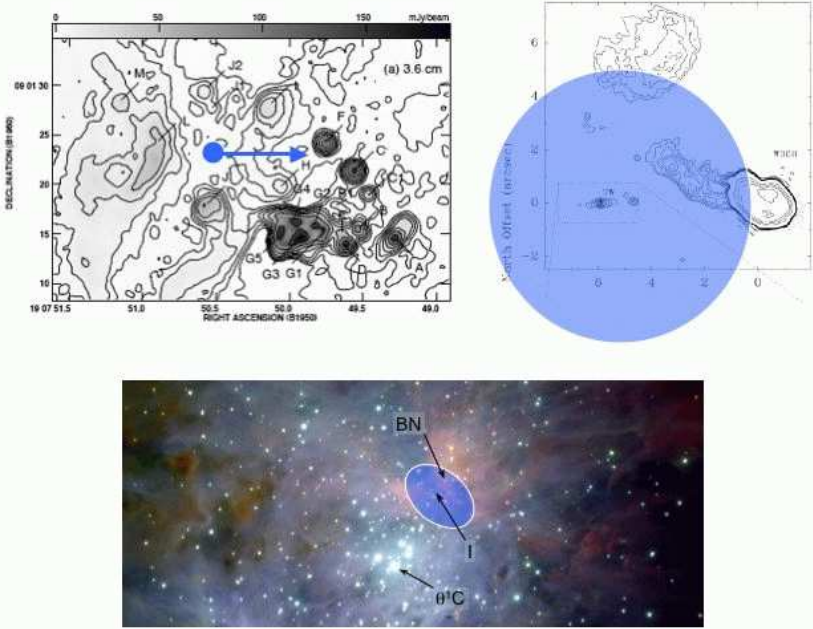}
  \caption{Comparison of the Turbulent Core Model (MT03) for the
  initial conditions of a $30\:M_\odot$ star, which forms from a
  $60\:M_\odot$ core, with three well-studied regions of massive star
  formation. The mean core radius is $R_{\rm core} \simeq 0.06
  (M_{\rm core}/60M_\odot)^{1/2}\Sigma^{-1/2}\:{\rm pc}$. In all
  cases we set $\Sigma=1\:{\rm g\:cm^{-2}}$ (i.e. the
  core is in pressure equilibrium with the central regions of a clump
  with this mean surface density). The different regions are: (a) {\it
  top left} the Welch ring of ultracompact HII regions in W49A shown
  in 3.6~cm emission (De Pree et al. 1997) --- core location here is
  arbitrary, while the adjacent arrow shows the distance the core moves
  traveling at $\sim 5\:{\rm km\:s^{-1}}$ during its collapse
  time of $\sim 10^5\:{\rm yr}$; (b) {\it top right} W3($\rm H_2O$) and
  W3(OH) region viewed at 3.6~cm (Wilner et al. 1999) --- the
  core is centered around W3($\rm H_2O$), as suggested by the
  observations of Wyrowski et al. (1999); (c) {\it bottom} Orion
  Trapezium and BN-KL shown in the near IR (VLT-ESO \& M. McCaughrean)
  --- the core is centered on source {\it I} (Menten \& Reid 1995)
  and elongated slightly to match the orientation of the observed sub-mm core
  (Wright et al. 1992). }
\label{fig:2}
\end{figure}

We now compare the predicted properties of the turbulent core model
(MT03) to some observed regions of massive star formation. The most
basic comparison we can make is the size of the core with the scale of
the region (Fig.~2).

\subsection{W49A}

W49A, 11.2~kpc away, is one of the richest regions of massive star
formation. Many ultracompact HII regions are seen in radio continuum
maps and each one may be powered by a young OB star. In Fig.~2 we show
the size of a $60\:M_\odot$ core in pressure equilibrium with the
central region of a $\Sigma=1\:{\rm g\:cm^{-2}}$ protocluster clump
(adopted as a typical value --- see Fig.~1 --- and not based on a particular
measurement of W49A). Fig.~2 also shows how far a core might move
during its star formation timescale, $\sim 10^5\:{\rm yr}$, if
traveling at the velocity dispersion of the system, taken to be $\sim
5\:{\rm km\:s^{-1}}$. We see that, while the region has a high density
of massive stars, this density does not prevent massive stars from
forming in relative isolation from each other, particularly allowing
for projection effects. Of course interactions with lower-mass stars
will be more common, but most of these would make only minor
perturbations to the structure of the core.

\subsection{W3($\rm H_2O$)}

W3($\rm H_2O$) and W3(OH) are much closer (2.07~kpc, Hachisuka et
al. 2004). While W3(OH) is an ultracompact HII region, probably
powered by a massive star that has completed its formation, W3($\rm
H_2O$) is likely to be a massive protostar, i.e. still growing
substantially by accretion. The luminosity of W3($\rm H_2O$) is $\sim
1-9\times 10^4 L_\odot$ (Wyrowski et al. 1999). MT03 used this to
estimate a current protostellar mass in the range $10 - 23
M_\odot$. Given that the star is still accreting and that Wyrowski et
al. (1999) estimate a gas mass of $\sim 15\:M_\odot$ for the inner
0.01~pc by 0.02~pc region, we again show the scale of a $60\:M_\odot$
core that might eventually form a $30\:M_\odot$ star. This scale is
several times larger than the detected 1.3~mm emission and a few times
larger than the $\rm C^{18}O$ emission (Wyrowski et al. 1999). To
reconcile this, one could consider somewhat lower mass cores in higher
pressure environments. However, note that the cores shown in Fig.~2
represent the initial configuration before onset of collapse. At later
stages the core becomes more concentrated. Also, there may be more
core material beyond the outer contour of $\rm C^{18}O$ emission shown
by Wyrowski et al. (1999). We conclude that the turbulent core model
works reasonably well in W3($\rm H_2O$). There are two radio continuum
sources in the core separated by just over 0.01~pc. If these are
powered by different stars and neglecting the possibility of
projection effects, then two stars are forming from the same core.

\subsection{Orion KL}

The closest massive protostar is thought to be radio source {\it I}
(Menten \& Reid 1995) in the Orion Kleinmann-Low (KL) region, $\sim
450$~pc away. This region is near the center of the Orion Nebula
Cluster (ONC), marked by the Trapezium OB stars (Fig.~2). Also nearby
is the Becklin-Neugebauer (BN) object, known to have a high proper
motion, equivalent to about 40~$\rm km\:s^{-1}$ in the plane of the sky
(Plambeck et al. 1995).
%Tan 2004c; Rodriguez et al. 2005).

\subsubsection{The Ejection of BN}

BN's luminosity is $\sim 2500-10^4\:L_\odot$ (Gezari, Backman, \&
Werner 1998), corresponding to a zero age main sequence B3-B4
(8-12~$M_\odot$) star. It is highly likely that BN originated in the
ONC. Since the cluster is too young for binary supernova ejections,
the most plausible model for BN's motion is dynamical ejection from an
unstable triple or higher order system. This can often occur when a
hard binary interacts with another star (Hut \& Bahcall
1983). Typically the least massive star is ejected at about the escape
speed from the remaining binary at the orbit of the secondary, which
is often left eccentric.

We have proposed BN was ejected from an interaction with the
$\theta^1C$ system (Tan 2004c) because: (1) $\theta^1C$ lies along
BN's past trajectory; (2) $\theta^1C$ has a proper motion direction
opposite to BN's (van Altena et al. 1988); (3) $\theta^1C$ has a
proper motion amplitude that would predict BN's mass is $6.4\pm3
M_\odot$, in agreement with the estimate from its luminosity;
%(the reason for the large error here is that we have allowed for a
%$\pm0.7$~mas/yr uncertainty in the initial motion of the system, which
%is the velocity dispersion of bright ONC stars (van Altena et
%al. 1988) and corresponds to $1.5\:{\rm km\:s^{-1}}$; 
%the estimate
%from the bolometric luminosity also has large uncertainties, extending
%beyond the mass range 8-12~$M_\odot$ that comes from assumption the
%star is on the zero age main sequence); 
(4) $\theta^1C$ has a relatively massive ($\gtrsim 6\:M_\odot$)
secondary companion (Schertl et al. 2003); (5) the projected
separation of the $\theta^1C$ binary (total mass $\simeq 50\:M_\odot$)
is 17AU and the escape speed from this distance is $70\:{\rm
  km\:s^{-1}}$, high enough to explain BN's speed. To have all of
the above occur by chance is highly improbable. Furthermore, no other
revealed, massive ONC stars have any of the correct proper motion or
binary properties.

Rodriguez et al. (2005, hereafter R05) measured proper motions of BN
and source {\it I} relative to a quasar $1.6^\circ$ away. Previous
measurements were of BN relative to source {\it I}, with the latter
assumed to be a massive protostar with a motion typical of the ONC.
R05 confirm the relative motion of BN and source {\it I}, but find
that source {\it I} is moving at $5.6\pm0.7$~mas/yr ($12\pm2\:{\rm
km\:s^{-1}}$) towards P.A.$+141^\circ \pm7^\circ$, i.e. away from
BN. R05 argue that source {\it I} ejected BN, which creates problems
for the interpretation that source {\it I} is a massive protostar
forming from an accretion disk (see below). However, source {\it I}'s
motion is roughly parallel to the Galactic plane and such
motions would not be unreasonable for the entire ONC due to Galactic
differential rotation. R05 also measure motions of 4 other nearby
radio sources thought to be ONC members, reporting motions
$\lesssim$3~mas/yr relative to the quasar.  However, these sources are
weak and undetected at some epochs. From deviations from straight
line motion we estimate positional uncertainties of $\gtrsim50$~mas in
some sources, leading to motion errors of $\gtrsim$4~mas/yr. We
conclude the data are not yet good enough to tell if source {\it I}
has a motion atypical of the ONC.

\subsubsection{Source I: Core, Disk, Protostar and Outflow}

%Core - wright et al. 92. Werner, Dinnerstein, Capps...

Wright et al. (1992) mapped a core of dense gas in the KL region in
emission at 450~$\rm \mu m$. This core is about a factor of two
smaller than that shown in Fig.~2, as might result from the
contraction that occurs after the onset of collapse. The polarization
vectors of near IR emission suggest that a single source is
responsible for much of the luminosity from the core (Werner, Capps,
\& Dinerstein 1983), and this source is likely to be the thermal radio
source {\it I} (Menten \& Reid 1995), which lies near the center of the
core.

%Disk - Wright et al. Greenhill et al.
Wright et al. (1995), Greenhill et al. (1998) and Tan (2004c) have
interpreted the system as containing a $r\sim 1000$~AU accretion disk,
as traced by SiO (v=0; J=2-1) maser emission, centered about source
{\it I}. This scale is similar to that estimated from collapse of a
core with $\beta=0.02$ (\S3.3). The velocity of maser spots from
different sides of the disk suggest a central mass of about
20~$M_\odot$. The disk alignment is perpendicular to the large scale
molecular outflow to the NW and SE (Chernin \& Wright 1996).

%Protostar - McKee & Tan - ion luminosity.

The luminosity of source {\it I} (we equate this to the region
often referred to as the ``Orion hot core'') is $1.3-5\times 10^4
L_\odot$ (Kaufman et al. 1998).
%, although
%could be as high as $10^5\:L_\odot$. 
MT03 used this to estimate a current protostellar mass of
$11 - 18 M_\odot$ and accretion rate
$2.2-5.0\times 10^{-4}\:M_\odot\:{\rm yr}^{-1}$.  The corresponding
H-ionizing photon luminosities are $10^{45} -
10^{48}\:s^{-1}$ (Tan \& McKee 2002).  
%The large range is because this
%is the mass range where the star undergoes rapid contraction towards
%the main sequence.

%Outflow - Tan & McKee 02. Chernin Wright. Dyson ....

Tan \& McKee (2003) modeled the HII region produced when an ionizing
source, i.e. a massive protostar, is embedded at the base of a neutral
disk wind or X-wind. This ``outflow-confined'', hyper-compact HII
region model appears to work well for source {\it I}: one model that can
explain the radio spectrum has an ionizing photon luminosity of
$2\times 10^{47}\:{\rm s^{-1}}$ and a mass outflow rate of $3\times
10^{-5}\:M_\odot\:{\rm yr^{-1}}$, about 10\% of the estimated
accretion rate. Source {\it I} is elongated (Menten \& Reid, in
prep.), and the position angle of elongation aligns well with a
Herbig-Haro object to the NW (Taylor et al. 1986) requiring flow
velocities $\sim 1000\:{\rm km\:s^{-1}}$, which is about the
escape speed from a 20~$M_\odot$ protostar. SiO (v=1 \& 2) masers have
been observed immediately surrounding the radio source on scales of
several tens of AU (Greenhill et al. 2003). The densities and
temperatures of the gas in the outflow-confined HII region model are
appropriate for the excitation of these masers. The velocities of
these masers are rather low ($\sim 10-20\:{\rm km\:s^{-1}}$) and
exhibit a gradient along the elogated direction of source {\it
I}. Greenhill et al. (2003) used this to argue that the disk is in
fact orientated along this axis, perpendicular to previous disk
models. However, there is little evidence for a large outflow in the
direction expected for this orientation and a new source would be
needed for the powerful NW-SE outflow.

In conclusion, many aspects of the Orion KL region and source {\it I}
can be explained by the Turbulent Core Model for massive star
formation. The scale of the core is relatively large, meaning that, in
the context of this model, it is not possible to form numerous massive
stars in this region. The assumption that the massive star forms in
isolation is obviously an approximation: the accretion flow can suffer
perturbations from the close passage of other stars, but only
relatively massive stars, such as BN, will have a potentially
significant effect.  Other stars are likely to be forming near the
collapsing massive core, but the opportunity for fragmentation from
the core itself is limited because it is globally collapsing on a
relatively short timescale, i.e. approximately its free-fall timescale.
The difference between this case and star clusters, which
do fragment into mostly low-mass stars, is probably the collapse time:
rich star clusters are likely to take many free-fall timescales to
form (Tan 2004a).

\begin{acknowledgments}
I thank many colleagues for discussions, particularly Chris McKee,
  Mark Krumholz, and Luis Rodriguez. 
%I thank Luiz Rodriguez for sharing proper motion data
% of stars in the Orion KL region.
 I am supported by a Zwicky fellowship from ETH.
\end{acknowledgments}

\end{document}